\documentclass[journal=jpclcd,manuscript=letter,layout=twocolumn]{achemso}
\usepackage[version=3]{mhchem} 
\usepackage{amsmath}
\usepackage{amssymb}
\usepackage{graphicx}
\usepackage{dcolumn}
\usepackage{bm}
\usepackage[colorlinks=true, allcolors=blue]{hyperref}

\setkeys{acs}{doi=true}

\newcommand{\doi}[1]{\href{https://doi.org/#1}{\textcolor{blue}{doi: #1}}} 

\author{Adrian Greichgauer} \email{greichgauer@ph2.uni-koeln.de}
\affiliation{Physics Institute II, University of Cologne, D-50937 K{\"o}ln, Germany}
\author{Roozbeh Yazdanpanah}
\affiliation{Physics Institute II, University of Cologne, D-50937 K{\"o}ln, Germany}
\author{Alexey Taskin}
\affiliation{Physics Institute II, University of Cologne, D-50937 K{\"o}ln, Germany}
\author{Oliver Breunig}
\affiliation{Physics Institute II, University of Cologne, D-50937 K{\"o}ln, Germany}
\author{Yoichi Ando}
\affiliation{Physics Institute II, University of Cologne, D-50937 K{\"o}ln, Germany}
\author{Jens Brede} \email{brede@ph2.uni-koeln.de}
\affiliation{Physics Institute II, University of Cologne, D-50937 K{\"o}ln, Germany}

\title{Ex Situ Fabrication of Superconducting Nanostructures for Low-Temperature STM}

\abbreviations{STM,STS,BST,TI,SC}

\begin{document}

\begin{tocentry}

\includegraphics[scale=1]{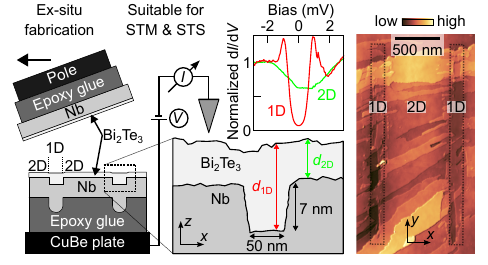}

\end{tocentry}

\begin{abstract}
Nanofabrication of topological insulator (TI) devices is essential for accessing edge and interface states, but conventional lithography and etching compromise the atomically clean surfaces required for scanning tunneling microscopy and spectroscopy (STM/STS). We demonstrate a fabrication strategy that combines ex situ lithographic patterning with in situ ultrahigh-vacuum cleaving and flipping, yielding atomically clean, nanopatterned TI–superconductor heterostructures suitable for STM/STS. In Design~I, nanoribbons were defined by etching trenches into a TI film and capping with Nb. This enabled spectroscopy on large areas, although edge quality was limited by etch debris. In Design~II, local thinning defined buried nanoribbons within a continuous TI film, producing pristine planar surfaces. STM/STS revealed well-developed superconducting gaps in the surrounding film, with suppressed gaps on the nanoribbons, consistent with vertical proximity coupling. This approach establishes a reproducible pathway for high-resolution STM/STS studies of proximitized nanostructures, providing a scalable platform for exploring topological superconductivity.
\end{abstract}


Topological insulator (TI) devices provide a versatile platform for studying emergent quantum phenomena, including topological superconductivity, the quantum anomalous Hall effect, and spin–orbit-driven transport~\cite{Breunig2022}. A key challenge is the limited reproducibility of such devices, often caused by material or fabrication-induced inhomogeneities. Random dopants and metal contacts shift the local chemical potential~\cite{Brede2024,Bai2020,Legg2022}, making spatially resolved STM/STS essential, but nanostructures from standard lithography rarely provide atomically clean surfaces.

Conventional nanofabrication leaves resist residues and surface damage that impair STM/STS~\cite{Ishigami2007}. Post-cleaning approaches (annealing, AFM cleaning~\cite{Goossens2012}) or surface protection by Se/Te capping~\cite{Virwani2014,Hoefer2015,Liang2021} either fail for fragile TIs~\cite{Luepke2020,Kong2025} or leave residues, and are difficult to integrate with lithographically patterned films.

A different route is to confine nanofabrication to one side of a heterostructure and expose a pristine surface by UHV cleaving. This “flip-chip” approach was pioneered by Stolyarov \emph{et al.}, who preserved lithographically patterned Nb hole arrays, but high-resolution STM was limited to unpatterned films~\cite{Stolyarov2014,Stolyarov2018}. 
Fl{\"o}totto \emph{et al.} later applied the method to TI–superconductor (SC) stacks, again only for unpatterned 2D films~\cite{Floetotto2018}. To date, STM/STS on cleaved, lithographically patterned TI-SC devices has not been achieved. However, such structures are indispensable for studying topological edge states that require well-defined nanoscale geometries. 

To address these limitations, we developed a fabrication strategy that combines ex situ lithography with in situ UHV cleaving, enabling atomically clean, nanopatterned TI--SC heterostructures suitable for high-resolution STM/STS. As a test case, we realized device geometries directly motivated by theoretical proposals.  

Design I is inspired by Cook and Franz~\cite{Cook2011}, who showed that a finite-length TI nanowire, proximity-coupled to an s-wave superconductor and subjected to a magnetic flux, can enter a one-dimensional topological superconducting phase hosting Majorana zero modes at its ends. To emulate this geometry, we sought to define quasi-one-dimensional nanoribbons within a continuous TI film and cap them with Nb. Such ribbons preserve both top and side TI surfaces, thereby realizing the geometry envisioned in Ref.~\cite{Cook2011}. 

Design II, following the proposal of Papaj and Fu~\cite{Papaj2021}, exploits thickness-dependent vertical proximity coupling: thinner regions acquire a strong induced gap, while thicker ribbons host low-energy modes that can become topological under an in-plane magnetic field. Our lithographic thinning approach implements this concept by embedding ribbons within a continuous TI film, avoiding etched sidewalls. In this way, Design II provides a direct experimental realization of the Papaj and Fu scenario, with quasi-one-dimensional channels confined by proximitized two-dimensional reservoirs. STM/STS measurements explicitly confirm the expected thickness-dependent gap contrast. While the present work does not target Majorana modes directly, it establishes control over the induced gap as a key design parameter for future studies of topological superconductivity in lithographically defined TI--SC nanostructures.

To implement these design concepts experimentally, we developed a fabrication process that evolved through three iterations, referred to as Generations~0, I, and II. Generations~0 and I are based on Design~I, while Generation~II corresponds to Design~II. Design~I features a rectangular array with a motif of two parallel lines that are etched away, leaving a nanoribbon defined in between. In contrast, Design~II also uses a rectangular array, but its motif consists of a single line that serves as a Nb etch mask, thereby defining the nanoribbon underneath.

\begin{figure}[h!]
	\centering
	\includegraphics[width=0.5\textwidth]{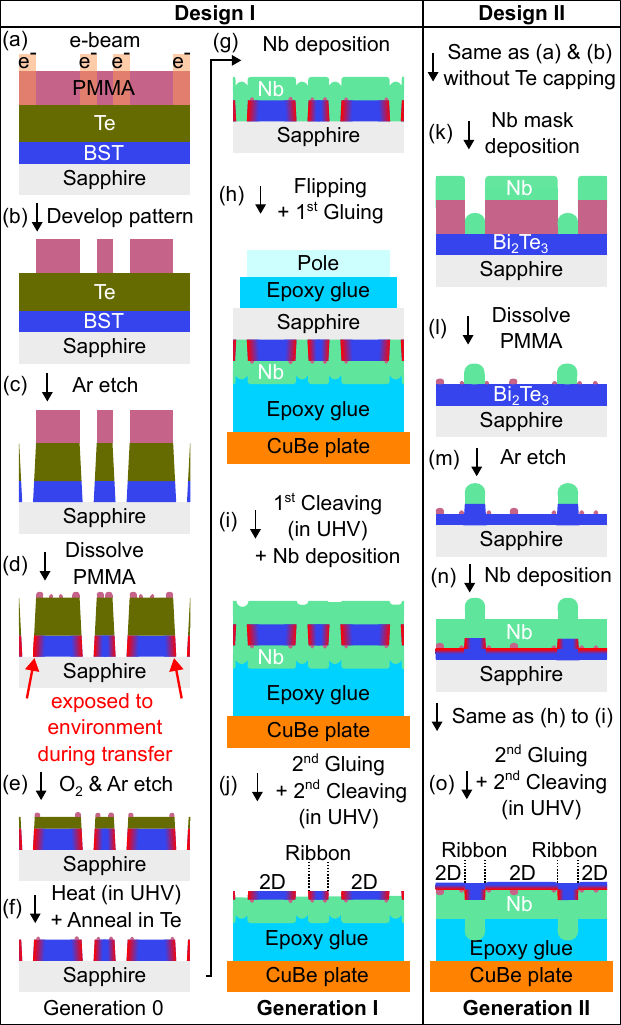}
	\caption{Fabrication schematics of our two designs leading to three sample generations. No STS on Generation 0; Results of Generation I and II shown in Figure~\ref{fig:GenerationI} and Figure~\ref{fig:GenerationII}.
    }
	\label{fig:Fabrication}
\end{figure}

For Design I we used (Bi$_{1-x}$Sb$_x$)$_2$Te$_3$ (BST) thin-films of $\sim$20\,nm thickness grown by molecular beam epitaxy (MBE) on sapphire substrates and capped with $\sim$75\,nm Te (unless otherwise stated) as described elsewhere~\cite{Yang2014,Taskin2017,Brede2024}. For Design II we used Bi$_2$Te$_3$ thin-films of the same thickness without Te capping\cite{Brede2024}.

The key fabrication steps are summarized in Figure~\ref{fig:Fabrication}\footnote{An extended schematic is provided in the Supporting Information Figure~S1,~S2.}.
Generation~0 employed Te-capped BST films patterned by e-beam lithography and etched down to the sapphire substrate. After resist removal, the samples underwent O$_2$/Ar plasma cleaning and thermal desorption of the Te cap under UHV with Te flux. 

Generation~I samples followed the same initial steps as Generation~0, after which a $\sim$50\,nm Nb layer was deposited in situ onto the patterned BST film. The stack was then flipped and cleaved under UHV to expose the opposite BST surface, which was largely unsuitable for STM/STS (Supporting Information Figure~S3). A second Nb deposition and UHV cleave yielded a significantly cleaner surface, as discussed later.

Generation~II samples implemented Design~II to avoid any direct nanofabrication on the STM-accessed surface. A stripe pattern was defined by e-beam lithography, Nb nanoribbons were formed to act as an etch mask after resist removal, and the Bi$_2$Te$_3$ between them was thinned by $\sim$7\,nm using Ar plasma etching. After deposition of a uniform Nb overlayer, the stack was flipped and cleaved twice under UHV. This placed the nanoribbons at the buried interface and left a pristine, continuous Bi$_2$Te$_3$ top surface ideally suited for STM/STS.

Having outlined the fabrication of our samples, we next examine their structural and spectroscopic properties. Due to residual contamination, the surface of Generation 0 (Supporting Information Figure~S4) was unsuited for spectroscopy, and we begin with Generation~I, where STM/STS provides direct insight into the surface quality and induced superconductivity.

\begin{figure*}[h!]
	\centering
	\includegraphics[width=\textwidth]{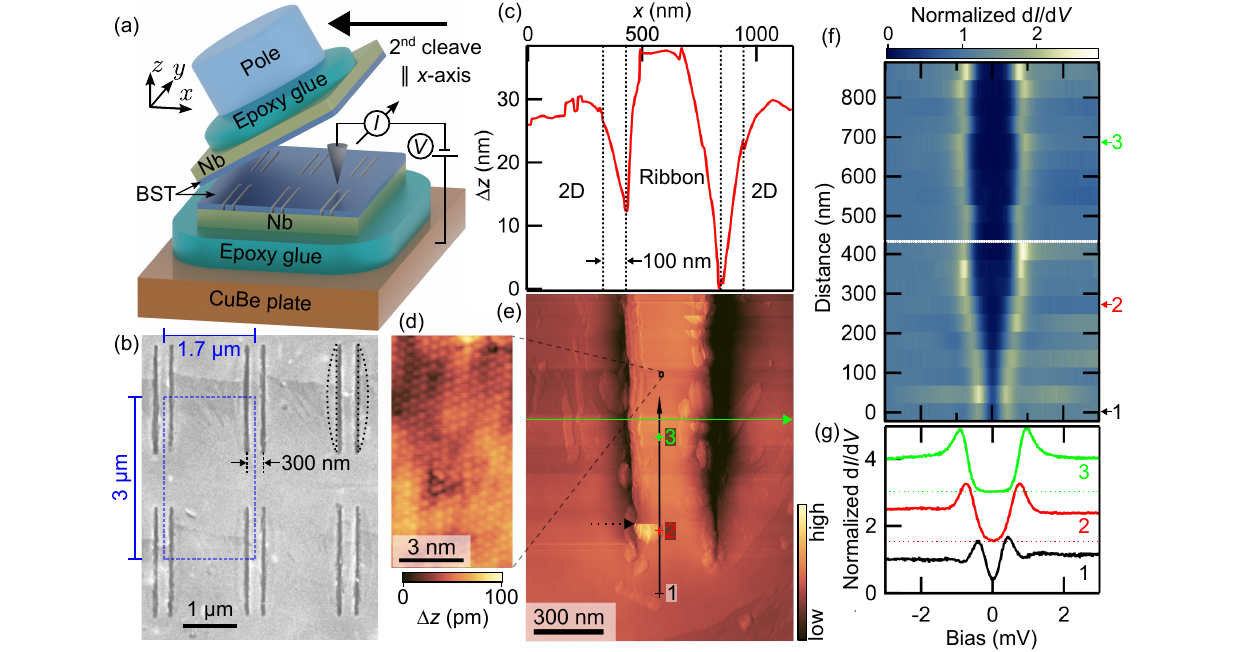}
    \caption{
    {\bf Nanoribbon array of a Generation I sample characterized by STM/STS.}
    (a) Schematic of the sample stack after UHV cleaving and in situ STM transfer. Bias is applied between the tip and CuBe plate.
    (b) SEM image of the nanoribbon array acquired post-STM; a sharpening filter highlights surface features.
    (c,e) STM topography (e) of a nanoribbon end; a gradient filter highlights edges (see SI S7). The green and black solid arrows mark the locations of the lineprofile (c) and STS line scan (f), respectively. Dotted arrow highlights nanoscale debris.
    (d) Atomically resolved image with drift correction~\cite{bagchi2025} applied.
    (f,g) Normalized differential conductance near the Fermi level. Spectra 1–3 in (g) correspond to traces in the line scan intensity plot (f), vertically offset for clarity. Corresponding locations are marked in (e). Dotted lines mark the zero d$I$/d$V$ level.
	Setpoints: (c,e) $I_0=20$\,pA, $V_0=990$\,mV; (d) $I_0=-0.5$\,nA, $V_0=-100$\,mV;
    (f,g) $I_0=0.5$\,nA, $V_0=10$\,mV, $V_\text{mod}=50$\,µV. All STM data acquired at 0.4\,K.
    }
	\label{fig:GenerationI}
\end{figure*}

The patterned array for Generation~I consisted of parallel trenches $\sim$300\,nm apart [Figure~\ref{fig:GenerationI}(a,b)]. STM topography reveals $\sim$50\,nm-wide, 20–30\,nm-deep trenches defining the ribbon, with flat terraces observed away from these features [Figure~\ref{fig:GenerationI}(c-e)]. Local bending near the ribbons likely originated from cleaving, but atomic-resolution images near the center of the ribbon confirmed an intact Te lattice with characteristic native defect signatures~\cite{Bagchi2022} [Figure~\ref{fig:GenerationI}(d)].

Spectra taken along a line that crosses the 2D film onto the ribbon [Figure~\ref{fig:GenerationI}(f)] evolved smoothly and were free of tip instabilities, confirming sufficient surface quality for spectroscopy\footnote{Supporting Information Figures~S5 and S6 provide complementary data on band structure, chemical potential, topography, and superconducting gap.}. The LDOS showed BCS-like gaps of $\sim$0.7\,meV on the ribbons and $\sim$0.2\,meV on the surrounding film [Figure~\ref{fig:GenerationI}(g)], indicating induced superconductivity in both regions and high-quality Nb/BST interfaces. Additional measurements on a second ribbon (Supporting Information Section~S5) revealed the opposite trend with smaller gaps on the ribbon than on the surrounding film. Line profiles of STS spectra across the nanoribbons (Supporting Information Figure~S6) showed no systematic evolution of the superconducting gap toward the etched sidewalls. Occasional variations were observed, but spectroscopy within a few lattice sites of the edges was hindered by tip instabilities from debris [Figure~\ref{fig:GenerationI}(e)]. Taken together, these results are consistent with a vertical proximity effect, where local thickness variations strongly influence the induced gap, as discussed below.

\begin{figure*}[h!]
	\centering
	\includegraphics[width=\textwidth]{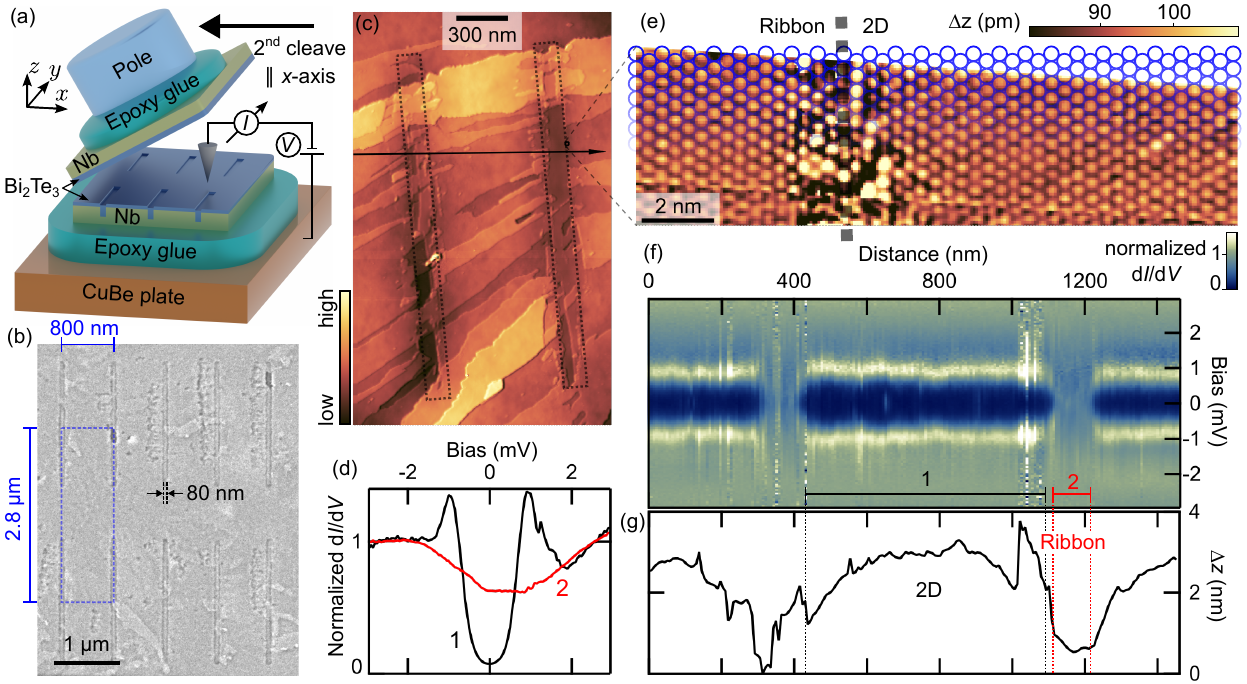}
	\caption{{\bf Nanoribbon array of a Generation II sample characterized by STM/STS.}
    (a) Schematic of the cleaved sample stack. Bias is applied between the tip and CuBe plate.
    (b) SEM image of the nanoribbon array post-STM; a sharpening filter highlights surface features.
    (c) Stitched STM topographies of two 
adjacent nanoribbons with a gradient filter to enhance edge contrast. Dotted rectangles outline nanoribbon positions; arrow indicates STS line in (f).
    (d) Averaged differential conductance spectra on (red) and between (black) nanoribbons, corresponding to intervals in (f).
    (e) Atomic-resolution image of the ribbon/film boundary with high-pass filtering and drift correction; blue circles mark an ideal hexagonal lattice, demonstrating lattice continuity.
    (f) Spatial map of normalized differential conductance at the Fermi level and corresponding height profile (g).
    Setpoints:
    (c) $I_0 = -30$\,pA, $V_0 = -900$\,mV;
    (d,f,g) $I_0 = -0.5$\,nA, $V_0 = -5$\,mV, $V_\text{mod} = 50$\,µV;
    (e) $I_0 = -0.5$\,nA, $V_0 = -5$\,mV.
    All STM data acquired at 0.4\,K.
    }
	\label{fig:GenerationII}
\end{figure*}

While Generation~I confirmed clean interfaces and robust induced superconductivity, it was not suitable for edge studies. Thus, we turned to Generation~II, where nanoribbons are defined by selective thinning rather than etching; the STM/STS characterization of such a sample is summarized in Figure~\ref{fig:GenerationII}.

The overview image in Figure~\ref{fig:GenerationII}(c), stitched from two STM scans, demonstrates that large areas can be imaged without scanning instabilities. Numerous elongated terraces are visible, roughly aligned along the $x$-direction, forming a tearing pattern that follows the cleaving direction [Figure~\ref{fig:GenerationII}(a)]. Interestingly, a slight curvature with minima $\sim$800\,nm apart is observed [Figure~\ref{fig:GenerationII}(g)]. Two faint $\sim$2~µm $\times$ 80\,nm structures can also be distinguished, consistent with the design shown in the SEM image [Figure~\ref{fig:GenerationII}(b)]. As expected for Design~II, no topographic discontinuity separates ribbon and film, since the ribbons are buried on the back side [Figure~\ref{fig:Fabrication}(o)]. Indeed, the atomic-resolution image [Figure~\ref{fig:GenerationII}(e)] at the ribbon/film boundary shows lattice continuity, with only minor distortions confined to a few sites.

This continuous planar interface, in contrast to the etched sidewalls of Generation~I, raises the question of whether quasi-one-dimensional ribbons are realized at all. The d$I$/d$V$ spectra near the Fermi level, taken along the path indicated by the arrow in Figure~\ref{fig:GenerationII}(c), address this point: regions between ribbons display a well-developed superconducting gap, while the nanoribbon regions exhibit significant LDOS at zero bias [Figure~\ref{fig:GenerationII}(f)]\footnote{Supporting Information Figure~S8 details the ribbon extent and additional STS data revealing the expected~\cite{Foerster2016} larger bulk band gap in the ultrathin 2D regions compared to $\sim$7\,nm thicker ribbons.}.

\begin{figure}[h!]
    \centering
    \includegraphics[width=0.5\textwidth]{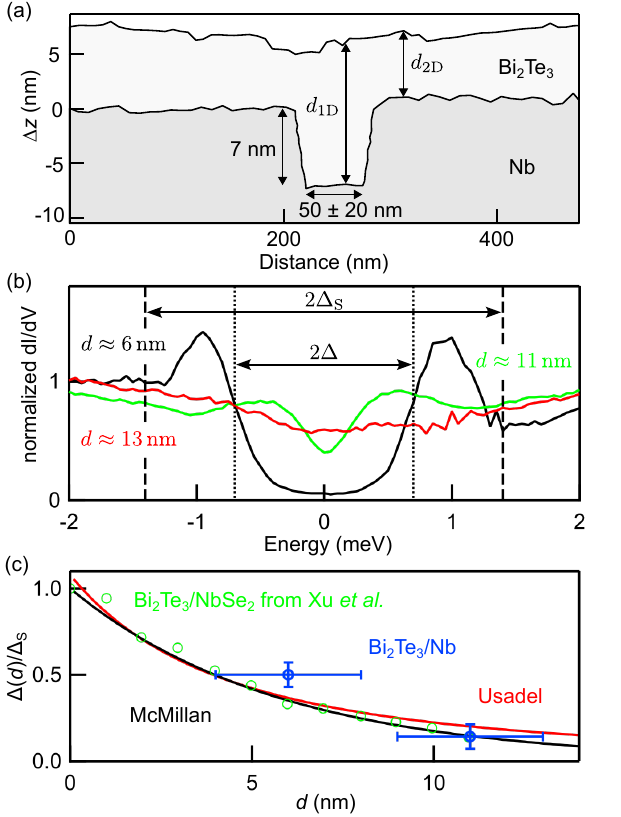}
    \caption{\textbf{Thickness dependence of the induced gap in Generation~II samples.}
    (a) Schematic cross-section illustrating TI thickness variations and surface roughness (cf.\ SI S9).
    (b) Representative spectra taken off a ribbon ($d\!\approx\!6$\,nm), on a ribbon ($d\!\approx\!13$\,nm), and at selected thinner ribbon location ($d\!\approx\!11$\,nm) [For STS positions see SI Figure~S8].
    (c) Induced gap $\Delta(d)$ compared with Bi$_2$Te$_3$/NbSe$_2$ data from Xu \emph{et al.}~\cite{Xu2014} and with diffusive (Usadel) and tunneling (McMillan) trends discussed in the text and SI Section~S7.}
    \label{fig:ABS}
\end{figure}

Average spectra for both regions are plotted in Figure~\ref{fig:GenerationII}(d). The details of the finite LDOS observed on the ribbons, requires further investigation, but it is clearly confined to the lithographically defined nanostructures.
We summarize the thickness-dependent variation of the measured spectra in Figure~\ref{fig:ABS} and discuss in the following that they are consistent with a vertical proximity effect, which produces a large induced gap in the surrounding 2D regions but only a weak or unresolved gap on the ribbons.

With a more detailed discussion of the different proximity models given in Supporting Information Section~S7, we summarize here only the key results: (i) the quasiclassical de Gennes–Saint-James (dGSJ) model~\cite{dGSJ1963} for a ballistic metal bounded by a superconductor and vacuum predicts negligible thickness dependence within our experimental range; (ii) native defects in the TI and interface roughness likely place the system in the diffusive regime, where the Usadel model~\cite{Usadel1970,Belzig1999} yields a stronger suppression of the induced gap $\Delta(d)$ with thickness [cf. Figure~\ref{fig:ABS}(c)], although reproducing the experiment would require an unrealistically short mean free path; (iii) the hybridization-limited McMillan model~\cite{McMillan1968}, $\Delta(d)\propto e^{-2\kappa d}$, provides a compact description consistent with microscopic Bogoliubov–de Gennes calculations~\cite{DasSarma2016}. Our representative $\Delta(d)$ values, together with Bi$_2$Te$_3$/NbSe$_2$ data by Xu \emph{et al.}~\cite{Xu2014}, are consistent with a rapid thickness-dependent suppression of the induced gap, as captured, for example, by the hybridization-limited vertical proximity model. However, we do not claim this as a unique explanation and note uncertainties in the absolute thickness calibration. Nonetheless, in the McMillan framework, local thickness variations of only a few nanometers already cause sizable changes in $\Delta(d)$, naturally explaining the gap inhomogeneity observed in Generation I. There, the stronger corrugation of BST films (Supporting Information Figure~S9) produced rougher Nb/TI interfaces, whereas the flatter Bi$_2$Te$_3$ films in Generation II enabled controlled thickness reduction and more uniform induced gaps.

In summary, samples from Generations 0 to II chart a clear progression.
Te-capped films (Generation 0) remained contaminated and unsuitable for STS.
Etched ribbons (Generation I) yielded clean Nb/TI interfaces and showed induced superconducting gaps but debris at etched sidewalls limited edge studies.
Selective thinning (Geneneration II) produced clean, continuous surfaces and a thickness-dependent induced gap consistent with a hybridization-limited vertical proximity effect and prior Bi$_2$Te$_3$/NbSe$_2$ results~\cite{Xu2014}, supporting Design II as a starting platform for realizing the Papaj and Fu proposal~\cite{Papaj2021}.

More broadly, the strong thickness dependence of the induced gap enables lithographic definition of proximitized reservoirs within metallic regions or metallic channels embedded in gapped superconducting backgrounds, providing a versatile platform for engineered proximity devices.

Two challenges remain for fabrication: the depth of the second cleave cannot be precisely controlled, and mechanical stress during cleaving can cause bending or tearing. Possible remedies include using alternative substrates such as InP (SI Fig.~S10) to eliminate the need for a second cleave, applying vertical rather than lateral force, and cooling the sample to stiffen the TI material to reduce stress. Design-based modifications such as auxiliary trenches to guide the peel front during cleavage, can help mitigate strain accumulation.

Beyond these technical points, our approach illustrates the broader advantages of MBE-grown films: periodic arrays with thousands of nanostructures, straightforward STM navigation without optical access, and compatiblity with surface-averaging techniques such as ARPES, which enables momentum-resolved characterization of nanopatterned systems~\cite{Mkhitaryan2024}. Together, these results establish a robust and broadly applicable fabrication pathway for atomically clean, lithographically defined TI-SC nanostructures, providing a versatile platform for STM/STS and complementary studies of proximitized topological states.

\begin{flushleft}
{\bf Data and materials availability:}\\
\end{flushleft}
\vspace{-3mm}
The data used in the generation of main and supporting figures are available from Zenodo\cite{Zenodo}.

\begin{acknowledgement}

This work has received funding from the Deutsche Forschungsgemeinschaft (DFG, German Research Foundation) under CRC 1238-277146847 (subprojects  A04, B01 and B06) as well as by the DFG under Germany’s Excellence Strategy -- Cluster of Excellence Matter and Light for Quantum Computing (ML4Q) EXC 2004/1-390534769. We also acknowledge the support of the DFG Major instrumentation program under project No. 544410649.

\end{acknowledgement}

\begin{suppinfo}

\section{Supporting Information}
Methods (fabrication process \& STM parameters, STM tip conditioning, data processing, volume of the study);
STM/SEM/AFM data of intermediate steps;
explanation of the gradient filter used for enhanced edge contrast;
additional data for Generation I (bulk bands);
second data set (different sample \& tip) for Generation I;
additional data for Generation II showing spatial extent of ribbons;
models for the vertical proximity effect;
fabrication on alternative InP substrate.

\end{suppinfo}


\providecommand{\noopsort}[1]{}\providecommand{\singleletter}[1]{#1}%
\providecommand{\latin}[1]{#1}
\makeatletter
\providecommand{\doi}
  {\begingroup\let\do\@makeother\dospecials
  \catcode`\{=1 \catcode`\}=2 \doi@aux}
\providecommand{\doi@aux}[1]{\endgroup\texttt{#1}}
\makeatother
\providecommand*\mcitethebibliography{\thebibliography}
\csname @ifundefined\endcsname{endmcitethebibliography}
  {\let\endmcitethebibliography\endthebibliography}{}

\end{document}


\section{Methods and Extended Fabrication}
\begin{figure}[htb]
	\centering
	\includegraphics[width=\textwidth]{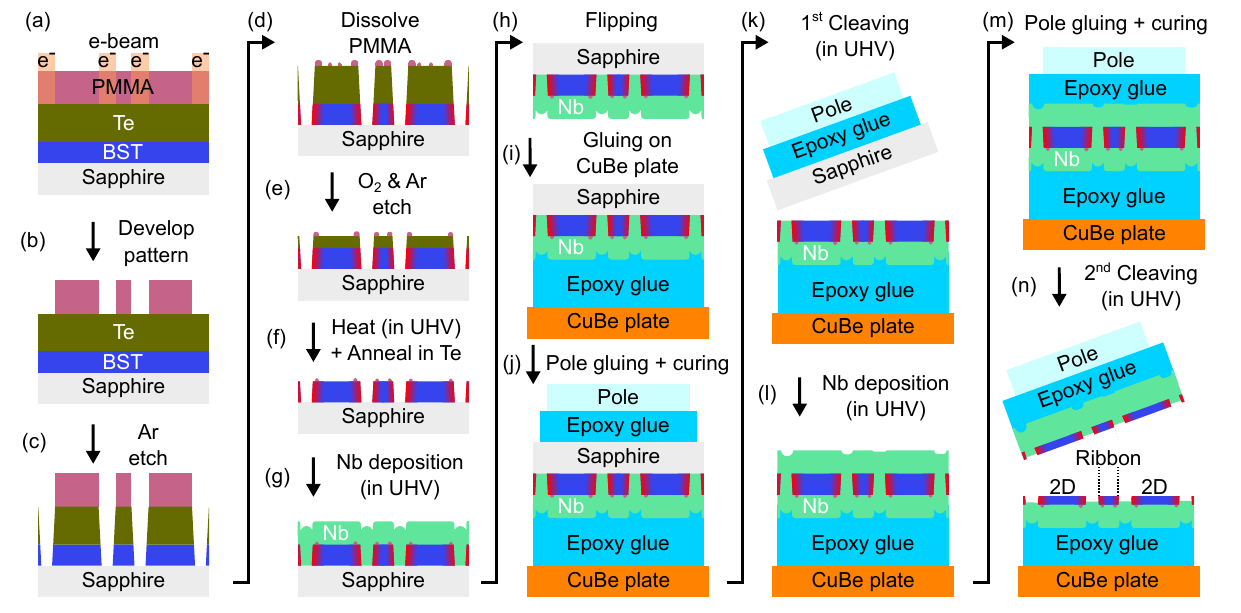}
	\vspace*{0mm}\caption{{\bf Full fabrication schematic of Design I.}
    Te-capped BST films were spin-coated with PMMA, patterned by e-beam lithography EBL (a), and developed in IPA/H\textsubscript{2}O (b). The pattern was etched into the BST layer via Ar plasma (c), and PMMA was removed with ultrasonic acetone/IPA baths (d). Residual contamination was minimized by sequential RIE with O\textsubscript{2} and Ar, followed by final Ar-ion sputtering in the STM prep chamber (e). Te decapping was done by annealing in UHV, with further annealing under Te flux (f). A Nb layer was then deposited in situ (g).
    After removing the sample from the UHV chamber, the structure was flipped (h), glued Nb-side-down to a CuBe plate (i), and a pole was attached to the sapphire side (j). Next, the sapphire was cleaved under UHV by applying a lateral force (k). A second Nb layer was deposited on the exposed BST surface (l), another pole was glued (m), and cleaved under UHV (n). Finally, the sample was transferred in situ to the STM for low-temperature measurements.
	}
	\label{fig:S1}
\end{figure}

\begin{figure}[htb]
	\centering
	\includegraphics[width=\textwidth]{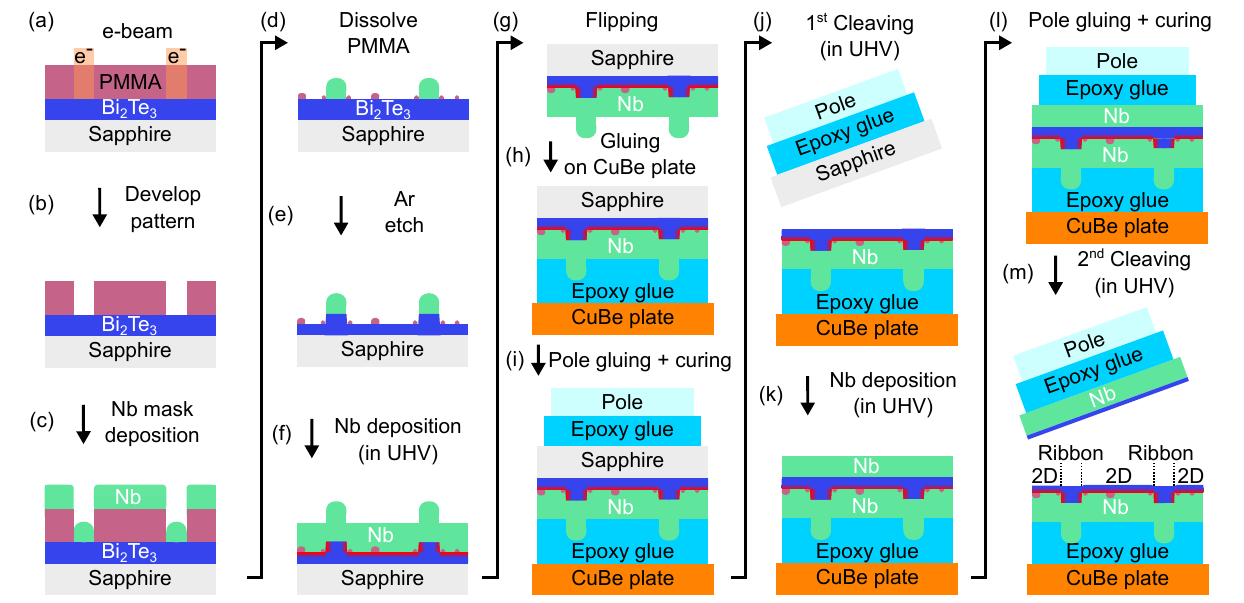}
	\vspace*{0mm}\caption{{\bf Full fabrication schematic of Design II.}
    Uncapped Bi\textsubscript{2}Te\textsubscript{3} films were spin-coated with PMMA and patterned via EBL (a), followed by development (b) and deposition of a Nb etch mask through the PMMA trenches (c). PMMA was removed by ultrasonic cleaning (d). Ar-based RIE thinned the Bi\textsubscript{2}Te\textsubscript{3} between ribbons (e), and a uniform Nb layer was deposited (f).
    The sample was then flipped (g), glued Nb-side-down to a CuBe plate (h), and a pole was attached to the sapphire side (i). The substrate was cleaved under UHV by lateral force (j). A second Nb layer was deposited on the exposed surface (k), a second pole was glued (l), and the stack was cleaved under UHV (m).
    Finally, the sample was transferred in situ to the STM chamber for low-temperature measurements.
	}
	\label{fig:S2}
\end{figure}

Key fabrication steps for Design I and II are shown in Figure~\ref{fig:S1} and Figure~\ref{fig:S2}, respectively. Extended schematics illustrate the full process flow, complementing the summary shown in Figure~1 of the main manuscript. In the following paragraphs, we discuss every process step starting from the molecular beam epitaxy (MBE) growth.


\paragraph{MBE growth:}
(Bi$_{1-x}$Sb$_x$)$_2$Te$_3$ (BST) and Bi$_2$Te$_3$ films with a thickness of about 20\,nm were grown on c-plane sapphire [$\text{Al}_2\text{O}_3$ (0001)] using the MBE technique as described elsewhere~\cite{Taskin2017, Yang2014}. After cooling to room temperature, the BST samples were capped with 75\,nm of Te to protect the films during ex-situ transfer and fabrication. Te-capped films had a surface roughness of about 4\,nm (RMS).
 
\paragraph{Fabrication Generation 0:}
The Te-capped BST films were removed from the MBE chamber and spin coated with about 300\,nm PMMA A4 resist. Nanoribbon patterns were defined using EBL (Raith Pioneer Two, $V_\text{acc} = 30$\,kV), followed by development in a 7:3 isopropanol/water (IPA/$\text{H}_2$O) solution. The pattern was transferred into the BST film via Ar plasma etching (Oxford Instruments PlasmaPro 80 reactive ion etch: $125$\,s, $200$\,W, $40$\,mTorr, 50\,sccm) and the best etch results were achieved for a total thickness of BST and Te of less than 100\,nm. PMMA was removed by successive ultrasonic baths in acetone and IPA at $50^\circ$C for 20\,min each.

To further reduce resist residues, the samples additionally etched with $\text{O}_2$-plasma in two steps: for $10$\,min with $20$\,W, $50$\,mTorr, 50\,sccm; and $2$\,min with $200$\,W, $40$\,mTorr, 50\,sccm, followed by Ar plasma etching ($20$\,s, $200$\,W, $50$\,mTorr, 50\,sccm). Final Ar-ion sputtering was performed in the STM preparation chamber ($4$\,min, $V_\text{acc} = 0.75$\,kV, $I_\text{em} = 7$\,mA, $p_\text{Ar} = 2 \times 10^{-6}$\,mbar).

Decapping was performed by annealing at 270$^\circ$C for 10\,min, followed by further annealing under Te flux at the same temperature for 150\,min. The sample was then transferred in situ to the STM measurement chamber, cooled, and subsequently characterized by STM.

\paragraph{Fabrication Generation I:}
The fabrication process continues with the samples from Generation 0. Without breaking vacuum, a Nb film ($\sim$50\,nm) was deposited on the patterned BST surface (Oxford Applied Research mini e-beam evaporator EGN4: $40$\,min, $70$\,W, $5 \times 10^{-10}$\,mbar).
After removing the sample from the STM preparation chamber, the Nb-coated surface was glued to a CuBe plate, and an Al pole was attached to the sapphire side using a two-component silver epoxy (EPO-TEK® H20E, curing: 30\,min, $120^\circ$C). The sapphire was then cleaved off under UHV conditions by applying lateral force via a UHV manipulator.
Immediately after cleaving, a second Nb layer ($\sim$50\,nm) was deposited ($40$\,min, $70$\,W, $5 \times 10^{-10}$\,mbar) onto the newly exposed BST surface. A second cleaving step followed: an Al pole was glued to this Nb layer using the same epoxy, and the stack was reintroduced into the STM preparation chamber, where the pole was cleaved off under UHV conditions.
The sample was then transferred in situ to the STM measurement chamber, cooled, and subsequently characterized by STM.

\paragraph{Fabrication Generation II:}
Here, uncapped Bi$_2$Te$_3$ films [AFM in Figure~\ref{fig:S9}(b)] were used. The nanoribbon pattern was written via EBL as described for Generation~0. The Nb etch mask ($\sim$50\,nm) was deposited ($35$\,min, $70$\,W, $1 \times 10^{-9}$\,mbar) onto the exposed Bi$_2$Te$_3$ surface through the developed PMMA trenches. Solvent-based PMMA removal was applied as described above [SEM of mask in Figure~\ref{fig:S9}(d,e)].
Subsequent Ar plasma etching was performed to thin the Bi$_2$Te$_3$ layer in the regions between the ribbons ($90$\,s, $50$\,W, $200$\,mTorr, 50\,sccm), reducing its thickness by approximately 7\,nm. A uniform Nb film ($\sim$80\,nm) was deposited across the entire sample ($60$\,min, $70$\,W, $5 \times 10^{-10}$\,mbar).
The sample was subsequently flipped, glued, cleaved twice, and transferred in situ into the STM measurement chamber, following the same procedure used for Generation I samples. The thicker Nb layers ($\sim$80\,nm) were used to enhance mechanical stability during cleaving and increase the reliability for successful cleaves.


\subsection{Scanning Probe Data Acquisition Parameters and Data Processing}

\paragraph{STM/STS:}
All STM experiments were performed using a commercial low-temperature UHV system (Unisoku USM1300) at 0.4\,K (unless stated otherwise). Topographic images were acquired in constant-current mode. For spectroscopy, the feedback loop was disabled after stabilizing at a chosen setpoint, and the bias voltage was ramped. Differential conductance (d$I$/d$V$) spectra were recorded using a lock-in amplifier with a small modulation voltage $V_\text{mod}$ at a frequency of 607\,Hz superimposed on the sample bias $V$.
Spectra of the SC gap were normalized by the smooth normal-state background, i.e., normalized $\mathrm{d}I/\mathrm{d}V \approx (\mathrm{d}I/\mathrm{d}V)_{\text{SC}} / (\mathrm{d}I/\mathrm{d}V)_{\text{normal}}$.
We used commercial platinum tips and in-house etched tungsten tips, which were cleaned either by a combination of Ar-ion sputtering ($30$\,min, $V_\text{acc} = 3$\,kV, $I_\text{em} = 7$\,mA, $p_\text{Ar} = 2 \times 10^{-6}$\,mbar) and short electron bombardment heating cycles (for $<10$\,s, $P_\text{tip} \approx 23$\,W), or by a single Ar-ion sputtering step ($70$\,min, $V_\text{acc} = 1.5$\,kV, $I_\text{fil} = 7$\,mA, $p_\text{Ar} = 2 \times 10^{-5}$\,mbar) with an applied tip bias of 150\,V~\cite{Schmucker2012}. Final tip conditioning was performed on a Cu(111) surface until clean surface state spectra were obtained.

\paragraph{AFM/SEM:}
All AFM measurements were performed at room temperature with two commercial systems (Park Systems NX10, Shimadzu SPM-9700). All SEM measurements were performed with our EBL system (Raith Pioneer Two).

\paragraph{Data processing:}
For all STM data from Design~I (except in Figure~\ref{fig:S3}), a piezo-recalibration factor of 1/1.3 was applied in the $x$-direction, either during acquisition or in subsequent analysis. All data was analyzed with {\small IGOR PRO} (Wavemetrics). The AFM data was processed with WSxM and Gwyddion.

\subsection{Volume of the study:}
In total, we measured 33 samples (excluding failed cleaves): Generation 0 ($N = 8$), Generation I ($N = 20$; 12 cleaved from sapphire, 8 from InP), and Generation II ($N = 5$). Cleaving success rates: 80\% (Generation I, sapphire), 65\% (Generation I, InP) and 100\% (Generation II, thicker Nb films). STM/STS acquired: Generation I ($N = 5$; 3 sapphire, 2 InP), and Generation II ($N = 2$). The remaining successful cleaves yielded only topographic data.

\newpage
\section{Intermediate steps of Design I}
\begin{figure}[H]
	\centering
	\includegraphics[width=\textwidth]{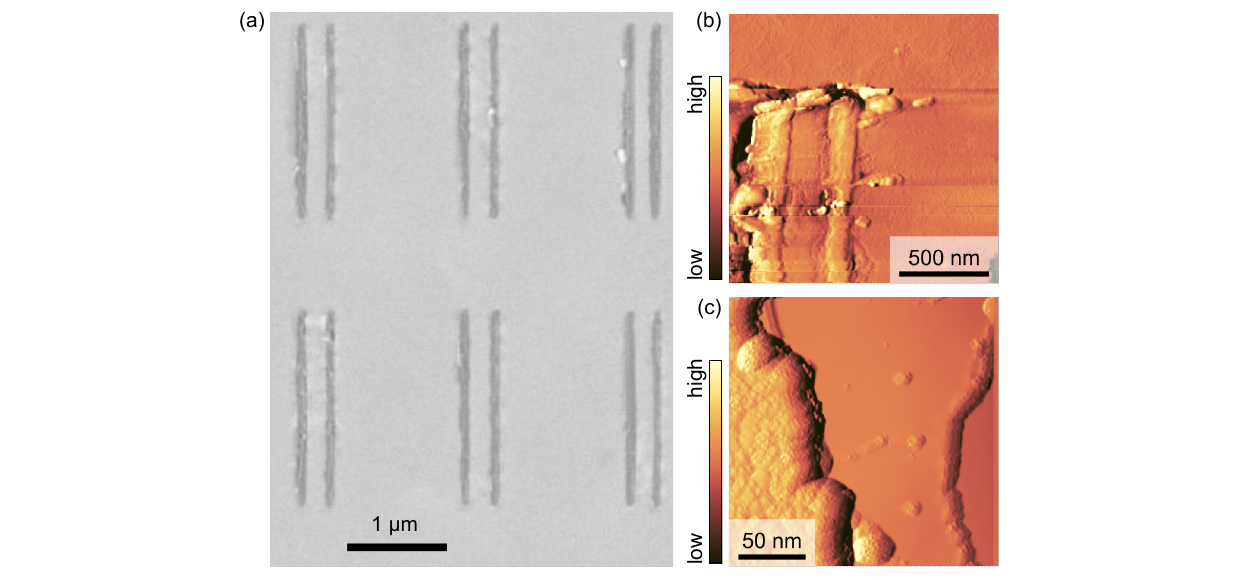}
	\vspace*{0mm}\caption{{\bf Characterization of BST surface after first cleaving from sapphire substrate.}
		(a) SEM image of the nanoribbon array on BST surface post-STM.
		(b) STM topography of the end of a fabricated nanoribbon structure. 
		(c) STM topography shows the difference between a surface cleaved in the wetting layer and an atomically flat cleave in the BST film. Only the atomically flat surface can be used to take meaningful STS data. A gradient filter highlights edges in (b,c).
		Setpoints: (b,c) $I_0=20$\,pA, $V_0=2$\,V.
		All STM data acquired at 0.4\,K.
		}
	\label{fig:S3}
\end{figure}

\begin{figure}[H]
	\centering
	\includegraphics[width=\textwidth]{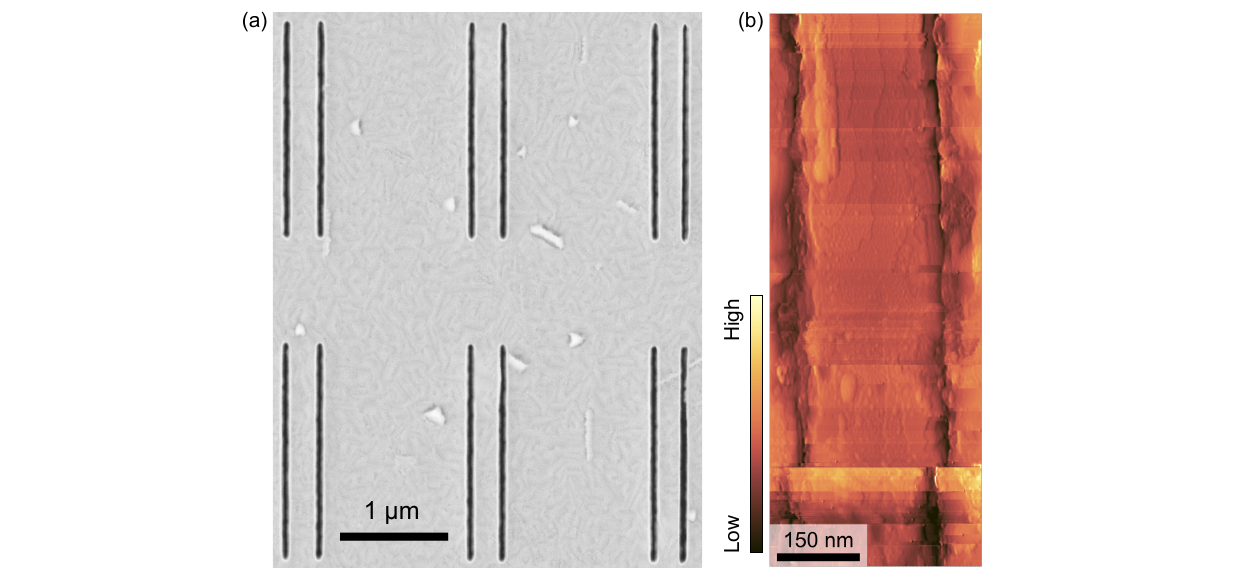}
	\vspace*{0mm}\caption{{\bf Characterization of BST surface before and after Te decapping.}
    (a) SEM image of a nanoribbon array on a Te-capped BST surface prior to STM measurements.
    (b) STM topography of a nanoribbon after Te decapping on the same sample shown in (a); a gradient filter enhances edge visibility.
    Data correspond to the same sample discussed in Figure~2 of the main manuscript and in Figures~\ref{fig:S7} and \ref{fig:S6}.
    Setpoint: (b) $I_0 = -20$~pA, $V_0 = -4$~V. STM measurements performed at 0.4\,K.
	}
	\label{fig:S4}
\end{figure}

As part of Design I, Generation 0 samples showed clean structures in SEM (Figure~\ref{fig:S4}), but Te decapping left residues that hindered STM beyond basic topography. This limitation led to Generation I, using a flip-chip approach. However, after the first UHV cleave, the surface was largely covered by a wetting layer, again unsuitable for STS (Figure~\ref{fig:S3}).

\section{Second Ribbon of Generation I sample}
\begin{figure}[H]
	\centering
	\includegraphics[width=\textwidth]{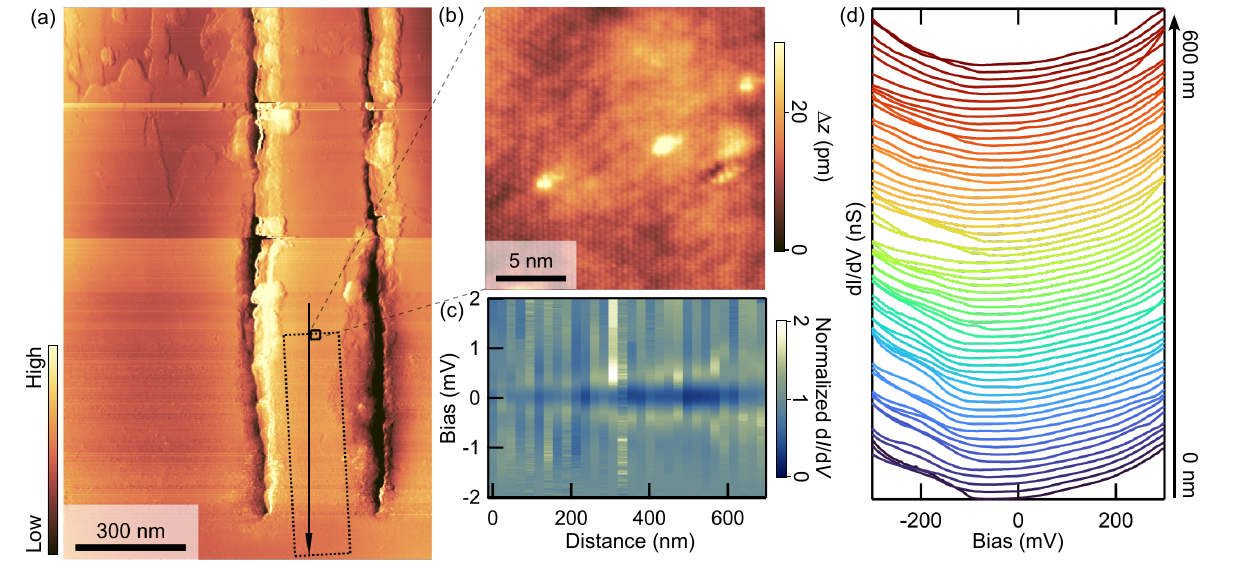}
	\vspace*{0mm}\caption{{\bf STM topography and STS characterization of a Generation I BST sample measured with a different tip, similar to Figure~2 in manuscript.}
    (a) STM topography of a nanoribbon end; a gradient filter highlights edges. The black arrow and dashed rectangle mark the locations of the STS line scan (c) and the $15 \times 60$ STS grid (150\,nm $\times$ 600\,nm) (d), respectively.
    (b) Atomically resolved image near the nanoribbon center (no drift correction or filtering applied).
    (c) Normalized differential conductance near the Fermi level showing the spatial evolution of a weakly induced SC gap.
    (d) Row-wise averaged d$I$/d$V$ spectra, vertically offset for clarity. Colors distinguish individual spectra.
	Setpoints: (a) $I_0=-20$\,pA, $V_0=-900$\,mV; (b) $I_0=-2$\,nA, $V_0=-900$\,mV; (d) $I_0=-1$\,nA, $V_0=-5$\,mV, $V_\text{mod}=50$\,µ$\text{V}$; (d) $I_0=1$\,nA, $V_0=300$\,mV, $V_\text{mod}=10$\,m$\text{V}$, $B_y = 300\,$mT;
	All data acquired at 0.4\,K.
	}
	\label{fig:S5}
\end{figure}

Here, we present data from another Generation I sample (different tip) to demonstrate that our fabrication approach reproducibly yields samples suitable for STM/STS investigations. The STM topography [Figure~\ref{fig:S5}(a)] shows a nanoribbon of similar quality to that discussed in the manuscript. The Te lattice is well resolved near the ribbon center (b), although residual debris at the edges hinders full characterization. The row-wise averaged high-bias $\mathrm{d}I/\mathrm{d}V$ spectra of the STS grid [dashed rectangle in [Figure~\ref{fig:S5}(a)]] are shown in Figure~\ref{fig:S5}(d), indicating reduced spatial variation of the chemical potential. However, near the Fermi level, high-resolution STS along the arrow in Figure~\ref{fig:S5}(a) reveals only a small superconducting (SC) gap, suggesting a weaker SC proximity effect likely due to locally thicker BST.

\section{Extended Data for Generation I}
\begin{figure}[ht]
	\centering
	\includegraphics[width=\textwidth]{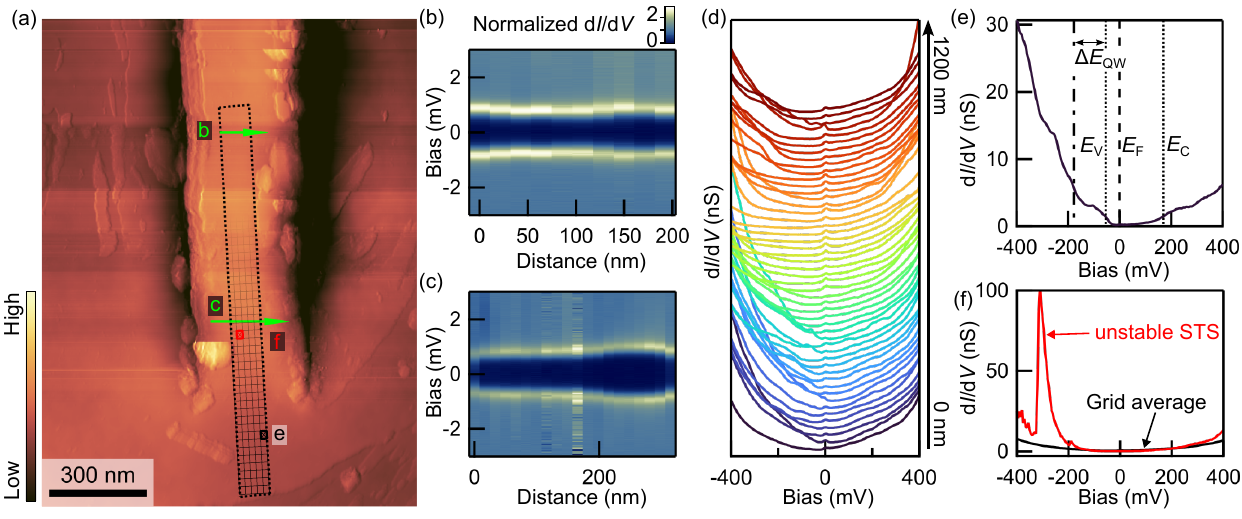}
	\vspace*{0mm}\caption{
	{\bf Additional data on the BST surface of the Generation I sample from Figure~2 in the main manuscript.}
    (a) STM topography of a nanoribbon end with gradient filter; dotted rectangle marks the $5\times50$ STS grid (120\,nm $\times$ 1200\,nm). Green lines indicate the line scan in (b,c); black/red rectangles mark the locations of spectra in (e,f).
    (b,c) Normalized d$I$/d$V$ near $E_F$ at positions indicated in (a).
    (d) Row-averaged d$I$/d$V$ spectra from the grid, excluding unstable spectra in (f); curves vertically offset, colors distinguish individual spectra.
    (e) Representative spectrum showing the local band structure of bulk-insulating BST.
    (f) Example of an unstable spectrum compared to the grid average.
    Setpoints:
	(a) $I_0=20$\,pA, $V_0=990$\,mV;
    (b,c) $I_0=500$\,pA, $V_0=10$\,mV, $V_\text{mod}=50$\,µV;
	(d,e,f) $I_0=0.5$\,nA, $V_0=300$\,mV, $V_\text{mod}=10$\,m$\text{V}$, $B_y=100$\,mT.
	All data acquired at 0.4\,K.}
	\label{fig:S6}
\end{figure}

In addition to the data shown in Figure~2 of the manuscript, we measured spatial variations of the SC gap across the ribbon and near its edges [Figure~\ref{fig:S6}(b,c)]. Although there are slight variations in the size of the gap, no systematic trend related to the edge is observed.
Furthermore, we characterized changes in the chemical potential by measuring the onset energies of the bulk bands~\cite{Brede2024}. An STS grid was acquired over the region marked by the dashed rectangle in Figure~\ref{fig:S6}(a). 
Figure~\ref{fig:S6}(d) shows row-wise averaged spectra from the grid, indicating notable lateral shifts in chemical potential along this nanoribbon.
A representative spectrum of the bulk-insulating BST is shown in Figure~\ref{fig:S6}(e), with the valence band edge $E_\text{V}$, conduction band edge $E_\text{C}$ and the energy difference between consecutive quantum well states $\Delta E_\text{QW} \approx 125$\,meV estimated from the first inflection point at energies below the valence band edge. The Fermi level $E_\text{F}$ is at zero bias. Notably, only one location in the grid exhibited an unstable tunnel junction, seen as an abrupt jump in the spectrum [red trace in Figure~\ref{fig:S6}(f)], located near nanoscale debris discussed in the manuscript.

\section{Gradient Filter for Enhanced Edge Contrast}

\begin{figure}[H]
	\centering
	\includegraphics[width=\textwidth]{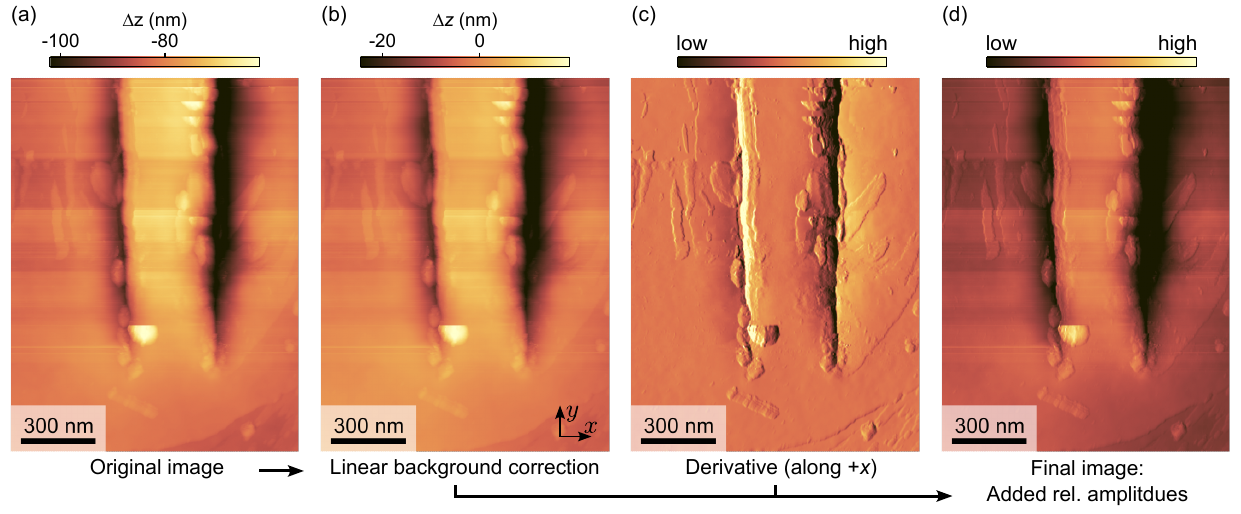}
	\vspace*{0mm}\caption{
	{\bf Gradient filtering of STM topography.}
    (a) Raw data.
    (b) Background-subtracted topography $z(x,y)$.
    (c) Horizontal gradient $\partial_xz(x, y)$.
    (d) Gradient-enhanced image: $z(x,y) + \mathcal{B}\partial_xz(x, y)$ with $\mathcal{B}=0.7$.
	}
	\label{fig:S7}
\end{figure}

To improve the visibility of nanostructure edges in STM topographies, we applied a gradient filter following the method described by Boshuis et al.~\cite{Bode2021}. Figure~\ref{fig:S7} illustrates the filtering process used throughout the manuscript.
Key features in each topography were enhanced using $z(x,y) + \mathcal{B}\partial_x z(x,y)$ with image-specific $\mathcal{B}$ values listed in Tab.~\ref{tab:B_value}.

\begin{table}[ht]
\centering
\caption{Summary of $\mathcal{B}$ values for the processed topography images}
\label{tab:B_value}
\begin{tabular}{lc}
\hline
\hline
Appearance & $\mathcal{B}$ value \\
\hline
Figure~2(e), Figure~\ref{fig:S7}(d), Figure~\ref{fig:S6}(a) & 0.7 \\
Figure~3(c), Figure~\ref{fig:S8}(a) & 0.3 \\
Figure~\ref{fig:S3}(b) & 1 \\
Figure~\ref{fig:S3}(c) & 10 \\
Figure~\ref{fig:S4}(b) & 0.7 \\
Figure~\ref{fig:S5}(a) & 1 \\
Figure~\ref{fig:S10}(h) & 0.5 \\
\hline
\hline
\end{tabular}
\end{table}

\section{Extended Data for Generation II}
\begin{figure}[H]
	\centering
	\includegraphics[width=\textwidth]{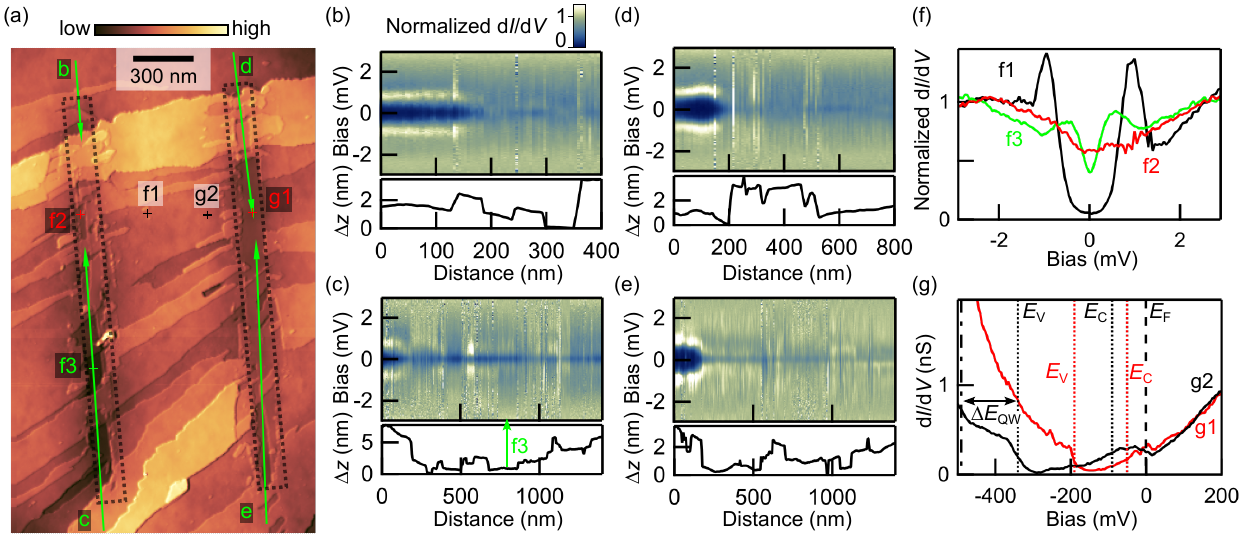}
	\vspace*{0mm}\caption{
	{\bf Additional data on Bi$_2$Te$_3$ surface of Generation II sample from Figure~3 in the main manuscript.}
    (a) Stitched STM topographies of adjacent nanoribbons with gradient filtering. Dotted rectangles mark ribbon        positions, green arrows indicate STS line scans in (b–e), and labeled crosses mark the spectra in (f,g).
    (b–e) Spatial maps of normalized d$I$/d$V$ at $E_F$ (top) with corresponding height profiles (bottom).
    (f) Representative d$I$/d$V$ spectra taken on nanoribbons (red, green) and in between (black), at positions indicated in (a,c).
    (g) d$I$/d$V$ spectra on (red) and between (black) nanoribbons.
	Setpoints:
	(a) $I_0 = -30$\,pA, $V_0 = -900$\,mV;
	(b-f) $I_0=-0.5$\,nA, $V_0=-5$\,mV, $V_\text{mod}=50$\,µV;
    (g) $I_0=0.5$\,nA, $V_0=400$\,mV, $V_\text{mod}=5$\,mV.
	All data acquired at 0.4\,K.
	}
	\label{fig:S8}
\end{figure}

In addition to the data shown in Figure~3 of the manuscript, large energy-range STS [Figure~\ref{fig:S8}(g)] shows a larger bulk-band gap off (black) than on (red) the nanoribbon, consistent with the increase in the bulk-band gap in ultrathin TI films\cite{Foerster2016}.

Additional STS across the ends of buried nanoribbons resolves the transition from a gapped 2D film to a quasi-1D ribbon [Figure~\ref{fig:S8}(b–e)]. At the lower end of the left ribbon [Figure~\ref{fig:S8}(c)], spectra taken on terraces $\sim$2 nm thinner than the surrounding ribbon show a small but finite superconducting gap. Spectra in Figure~\ref{fig:S8}(f) [cf. Figure~4(b)] were acquired at the sites marked in (a), while the large-bias spectrum in (g) reveals quantum well states with spacing $\Delta E_\text{QW}\approx 150$ meV, used to calibrate the local thickness of the 2D region ($d_\text{2D}\approx 6$ nm).

\section{Models for the Vertical Proximity Effect}

In the main text we summarize the thickness dependence of the induced gap and compare it with ballistic, diffusive, and hybridization-limited pictures. 
Here we provide the underlying model expressions and simple estimates that motivate those comparisons.

In the first approximation, the normal‐conducting TI film of thickness $d$ has no intrinsic superconducting order. Pairing correlations are induced by proximity to Nb. Restricting to ballistic trajectories at subgap energies $0<E<\Delta_\text{S}$ (with a Nb gap of our films of $\Delta_\text{S}\!\approx\!1.4$\,meV), the first de Gennes–Saint-James (dGSJ) bound state~\cite{dGSJ1963} $E_{n=0}$ arises from closed orbits made of Andreev reflection at the S–N interface and specular reflection at the TI–vacuum interface. Its quantization reads
\begin{equation}
    \frac{2 E_{0} d}{\hbar v_\text{F} \cos\theta}
    =\arccos\!\left(\frac{E_{0}}{\Delta_\text{S}}\right),
\end{equation}
with $v_\text{F}=\hbar k_\text{F}/m_{\mathrm{eff}}$, $m_{\mathrm{eff}}$ the effective mass, and $k_\perp=k_\text{F}\cos\theta$ the component normal to the interface. Since the STM tunneling matrix element favors near‐normal incidence, we set $\cos\theta=1$. Approximating the bulk conduction band as free‐electron–like with $m_{\mathrm{eff}}=0.3\,m_\text{e}$, we estimate $v_\text{F}=\sqrt{2E_\text{F}/m_{\mathrm{eff}}}$ from the experimental $E_\text{F}\!\approx\!75$~meV [Figure~S8(g)]. For $d=6$–13\,nm the lowest dGSJ level lies very close to $\Delta_\text{S}$ with negligible thickness dependence. The characteristic $1/d$ dispersion appears only beyond
\[
    d^\star \simeq \frac{\pi}{4}\,\frac{\hbar v_\text{F}}{\Delta_\text{S}}\approx 110\,\mathrm{nm}\quad(\Delta_\text{S}\approx 1.4\,\mathrm{meV}),
\]
so a purely ballistic dGSJ picture cannot explain the strong thickness contrast we observe (coherence-peak–like LDOS features at $d=6$\,nm that are absent at $d=13$\,nm).

Native defects in Bi$_2$Te$_3$ and roughness at the Nb–TI interface can mix momenta, plausibly placing the proximity effect in the diffusive regime. In that limit, solving the Usadel equations~\cite{Usadel1970,Belzig1999} with a transparent S boundary and a reflecting outer surface (S-N-I) gives
\begin{equation}
    \Delta(d)=\Delta_\text{S}\left(1+\alpha\,\frac{d}{\xi_\text{N}}\right)^{-2},
\end{equation}
where
\begin{align*}
    &\xi_\text{N} =\sqrt{\frac{\hbar D}{\Delta_\text{S}}}, 
    && D=\frac{v_\text{F} \ell}{3},
    &&\alpha=\frac{1}{\sqrt{c}} \approx \frac{1}{\sqrt{0.78}} \approx 1.13,
\end{align*}
so the thick‐film tail matches the S-N Usadel result $\Delta(d)\simeq c\,\hbar D/d^2$ with the universal S-N-I prefactor $c=0.78$. Using $v_\text{F}$ as above and the experimental values $\Delta_\text{S}\approx 1.4$~meV, $\Delta(d=6~\mathrm{nm})\approx 0.7$~meV, and $\Delta(d=13~\mathrm{nm})<0.15$~meV implies a mean free path $\ell\approx 0.6$\,nm. This violates the quasiclassical condition $k_\text{F}\ell\gg1$, indicating additional suppression beyond a bare diffusive minigap or the ballistic dGSJ picture.

A compact way to capture additional suppression is the tunneling (McMillan) viewpoint~\cite{McMillan1968}, where the induced scale is set by the hybridization rate $\Gamma$ into S, $\Delta(d)\!\approx\!\Gamma=\hbar/\tau_{\mathrm{esc}}$. For states localized at the free TI surface, any barrier or momentum selection across the film makes propagation through the TI evanescent. Writing the through–TI propagator as $G_\text{N}(0,d)\!\propto\!e^{-\kappa d}$ (with $\kappa$ the smallest imaginary longitudinal wave vector in the TI at $E_\text{F}$) gives
\[
    \Delta(d)\ \propto\ |t_{\mathrm{eff}}|^2\ \propto\ e^{-2\kappa d}.
\]
In particular, if the dominant channel is the free–surface topological surface state and no bulk propagating partner exists at the same $k_\parallel$, the coupling to S is exponentially weak and the free–surface minigap decays rapidly with $d$.

A fully microscopic description is provided by BdG calculations for TI/SC slabs by Chiu \emph{et al.}~\cite{DasSarma2016}, which show that the induced \emph{spectral} gap on the free TI surface decays with thickness as \(\Delta(d)\!\propto\!e^{-2\kappa d}\) (with oscillations) and, in the metallic regime, crosses over to an algebraic \(\Delta(d)\propto d^{-3}\) law. In their clean limit, the surface–state amplitude decay length is \(1/\kappa\simeq 0.875\)\,nm, corresponding to an energy decay length \(\Lambda_E\equiv 1/(2\kappa)\!<\!0.5\)\,nm. By contrast, the longer experimental \(\Lambda_E\sim 5\)–7\,nm reported for Bi$_2$Te$_3$/NbSe$_2$ by Xu \emph{et al.}~\cite{Xu2014} can be understood by a softened effective barrier in real films—electrostatic band bending and potential fluctuations shifting local band edges by tens of meV~\cite{Brede2024}—together with momentum mixing from defects and Nb–TI interface roughness, all of which reduce the effective \(\kappa\).

Taken together, these simple model estimates support the interpretation in the main manuscript: ballistic quantization produces little variation in our thickness range, diffusive models require unphysical parameters, and a hybridization-limited picture yields exponential suppression consistent with our Generation~II data and with Xu \emph{et al.}.

\section{Film Topography After MBE Growth and with Nb Mask}
\begin{figure}[H]
	\centering
	\includegraphics[width=\textwidth]{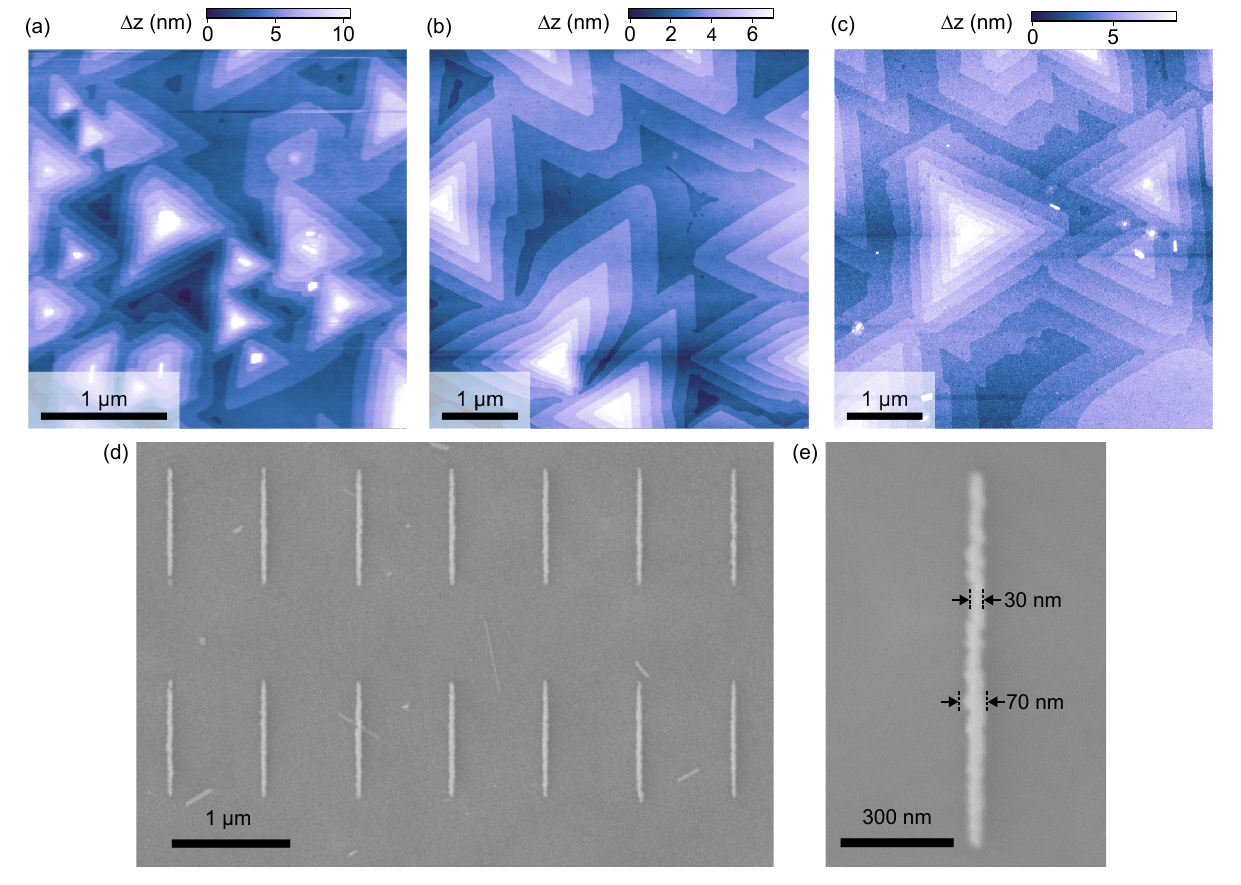}
	\vspace*{0mm}\caption{
    \textbf{Topography of BST (Generation I), and Bi$_2$Te$_3$ films together with Nb mask of Generation II samples.}
    (a–c) AFM images of (a) BST, (b) Bi$_2$Te$_3$, and (c) Bi$_2$Te$_3$ after Ar etching (150 s, 50 W, 200 mTorr, 50 sccm). Note the different scale bars in (a) vs. (b,c).
    (d,e) Nb mask of a Generation II sample after PMMA removal [cf. Figure~\ref{fig:S2}(d)], with minimum and maximum mask widths indicated.
	}
	\label{fig:S9}
\end{figure}

AFM measurements of representative as-grown BST [Figure~\ref{fig:S9}(a)] and Bi$_2$Te$_3$ films before [Figure~\ref{fig:S9}(b)] and after Ar etching [Figure~\ref{fig:S9}(c)] reveal distinct surface morphologies. BST films exhibit smaller triangular islands, whereas Bi$_2$Te$_3$ shows larger islands. Ar etching is minimally invasive: it preserves the island structure of Bi$_2$Te$_3$, removes the film layer by layer, and induces only slight terrace corrugation visible in [Figure~\ref{fig:S9}(c)].
SEM images of  Bi$_2$Te$_3$ films after Nb mask deposition [cf.\ Figure~\ref{fig:S2}(a-d)] for Generation II show a mask width of roughly $50\pm20$\,nm [Figure~\ref{fig:S9}(d,e)] 

\section{Fabrication on Alternative Substrate}
\begin{figure}[H]
	\centering
	\includegraphics[width=\textwidth]{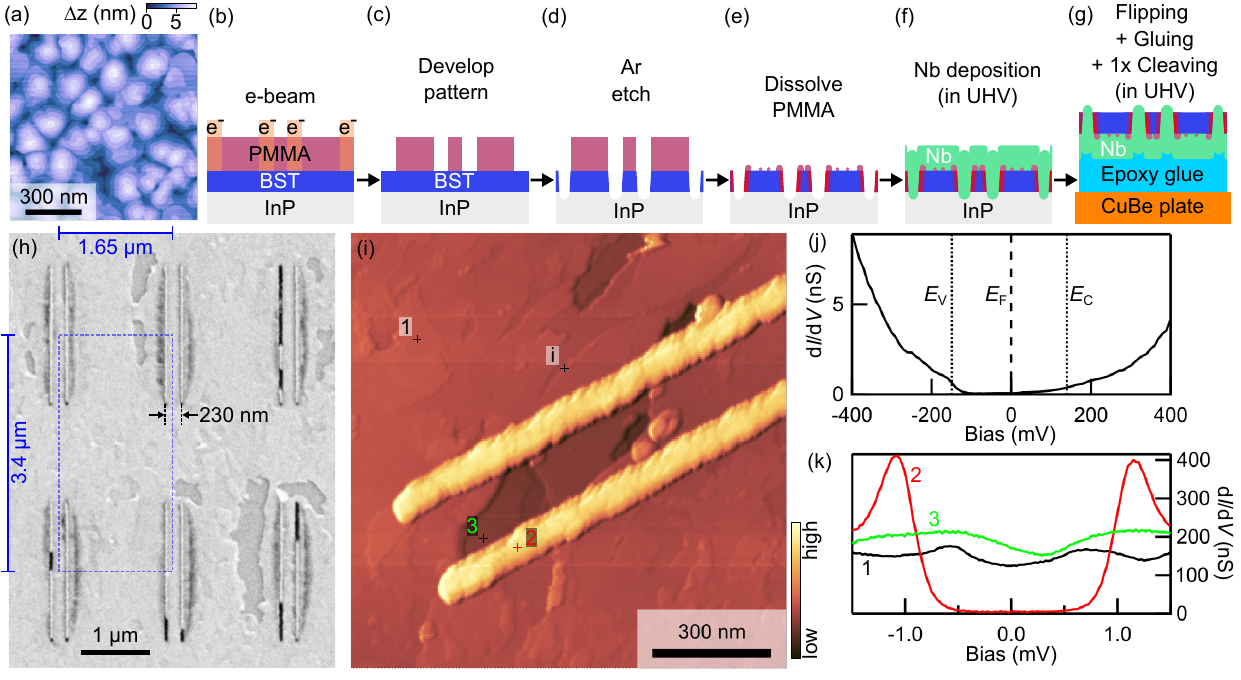}
	\vspace*{0mm}\caption{
    \textbf{Nanoribbon array of a Generation I sample cleaved from InP substrate and characterized by STM/STS.}
    (a) AFM topography of BST grown on InP(111).
    (b–g) Schematic of the fabrication procedure, following the Generation I steps: EBL (b,c), Ar etching (60\,s, 200\,W, 40\,mTorr, 50 sccm) through the BST (d), PMMA removal (e), Nb deposition (f), and UHV cleaving (g). Due to the lower etch resistance of InP, partial substrate etching during (d) leads to Nb ridges surrounding the nanoribbons (g), visible in SEM and STM (h,i).  
    (h) SEM image of the BST surface after STM.  
    (i) STM topography of a nanoribbon end with gradient filter; crosses mark positions of d$I$/d$V$ in (j,k).  
    (j) Single d$I$/d$V$ spectrum shows the local band structure of the bulk-insulating BST.
    (k) d$I$/d$V$ at $E_\text{F}$ in the 2D BST region (1), on the Nb ridge (2), and the nanoribbon (3).
	Setpoints:
	(i) $I_0 = -20$\,pA, $V_0 = -900$\,mV;
	(j) $I_0=0.5$\,nA, $V_0=400$\,mV, $V_\text{mod}=10$\,mV;
	(k) $I_0=-2$\,nA, $V_0=-10$\,mV, $V_\text{mod}=50$\,µV.
	Data in (j) was measured at 1.5\,K, other STM data acquired at 0.4\,K.
	}
	\label{fig:S10}
\end{figure}

Here, we present additional data showing that the fabrication recipe of Generation I samples is not limited to sapphire substrates. About 30\,nm thick BST films grown on InP(111) (no Te capping) also show promise (Figure~\ref{fig:S10}), as they do not exhibit the wetting layer commonly seen after the first cleave on sapphire-grown films (Figure~\ref{fig:S3}). Eliminating the need for a second cleave reduces mechanical strain during fabrication. Interestingly, Nb adhesion to InP is notably weaker, resulting in Nb ridges visible as parallel lines in the STM topography [Figure~\ref{fig:S10}(i)]. These ridges may be avoided by stopping the etch above the InP substrate [Figure~\ref{fig:S10}(d)] before Nb deposition. However, the current geometry offers an opportunity to study the lateral proximity effect between the Nb ridges and the adjacent BST [Figure~\ref{fig:S10}(k)].

\providecommand{\noopsort}[1]{}\providecommand{\singleletter}[1]{#1}%
\providecommand{\latin}[1]{#1}
\makeatletter
\providecommand{\doi}
  {\begingroup\let\do\@makeother\dospecials
  \catcode`\{=1 \catcode`\}=2 \doi@aux}
\providecommand{\doi@aux}[1]{\endgroup\texttt{#1}}
\makeatother
\providecommand*\mcitethebibliography{\thebibliography}
\csname @ifundefined\endcsname{endmcitethebibliography}
  {\let\endmcitethebibliography\endthebibliography}{}